\documentclass[usenatbib]{mn2e}
\usepackage{amssymb}
\usepackage{times}
\usepackage{epsfig}

\newcommand{\xte}{{\it RXTE}}
\newcommand{\gro}{{\it CGRO}}

\title[Superorbital variability of X-ray and radio emission of Cyg X-1]{Superorbital variability of X-ray 
and radio emission of Cyg X-1.\\ I. Emission anisotropy of precessing sources}

\author[A. Ibragimov, A. A. Zdziarski and J. Poutanen]
{Askar Ibragimov,$^{1,2}$\thanks{E-mail:
askar.ibragimov@oulu.fi (AI), aaz@camk.edu.pl (AAZ)} 
Andrzej A. Zdziarski$^3$\footnotemark[1] and 
Juri Poutanen$^1$\\
$^1$Astronomy Division, PO Box 3000, FIN-90014 University of Oulu, Finland\\
$^2$Kazan State University, Astronomy Department, Kremlyovskaya 18, 420008
Kazan, Russia\\
$^3$Centrum Astronomiczne im.\ M. Kopernika, Bartycka 18, 00-716 Warszawa, Poland\\
}

\date{Accepted 2007 July 11. Received 2007 June 05; in original form 2007 March 27}

\pagerange{\pageref{firstpage}--\pageref{lastpage}}
\pubyear{2007}

\begin{document}

\maketitle

\label{firstpage}

\begin{abstract}
We study theoretical interpretations of the $\sim\! 150$-d (superorbital) modulation observed in X-ray and radio emission of Cyg X-1 in the framework of models connecting this phenomenon to precession. Precession changes the orientation of the emission source (either disc or jet) relative to the observer. This leads to emission modulation due to an anisotropic emission pattern of the source or orientation-dependent amount of absorbing medium along the line of sight or both. We consider, in particular, anisotropy patterns of blackbody-type emission, thermal Comptonization in slab geometry, jet/outflow beaming, and absorption in a coronal-type medium above the disc. We then fit these models to the data from the \xte\/ All Sky Monitor, \gro\/ BATSE, and the Ryle and Green Bank radio telescopes, and find relatively small best-fit angles between the precession and orbital planes, $\sim\! 10\degr$--$20\degr$. The thermal Comptonization model for the X-ray emission explains well the observed decrease of the variability amplitude from 1 to 300 keV as a result of a reduced anisotropy of the emission due to multiple scatterings. Our modeling also yield the jet bulk velocity of $\sim (0.3$--$0.5)c$, which is in agreement with the previous constraint from the lack of an observed counterjet and lack of short-term X-ray/radio correlations. 
\end{abstract}

\section{Introduction}
\label{s:intro}

Periodic variability of emission from Cyg X-1 flux at various frequencies at the period of $\sim$150 d has been reported by, e.g., \citet*{brock}, \citet*{poo99}, \citet{od01}, \citet{benlloch01, benlloch04}, \citet{k01} and \citet[][ hereafter L06]{l06}. This period is much longer than the 5.6-d orbital period \citep{brock2}, and this type of periodicity (or quasi-periodicity) in binaries is called superorbital. The generally accepted interpretation of the underlying cause of superorbital periodicity in X-ray binaries is precession of the accretion disc and/or jet (e.g.\ \citealt{k73,k80,l98,wp99,ogdu01,tor05,cap06}; L06; but with the exception of 4U 1820--303, e.g., \citealt*{zwg07}). However, the question arises in which way the precession causes the modulation of the observed flux. There appears to be a number of possibilities. 

Considering the X-ray modulation first, the outer edge of the optically thick disc may partially cover the X-ray source. This, however, would require extreme fine-tuning. Namely, the X-ray source has the size $\sim 10^2 R_{\rm g}$ (where $R_{\rm g}\equiv GM/c^2$), as indicated by the X-ray power spectrum and agreement with theoretical prediction on the range of radii where most of the accretion power is released, while the disc size is generally much larger, up to the order of the size of the Roche lobe ($\sim 10^6 R_{\rm g}$ in Cyg X-1). Another possibility is that the outer part of the disc fully obscures the X-ray source, but we see the X-rays scattered in a large corona above the disc (this appears to take place, e.g., in Her X-1, \citealt{l02}). This, however, would dramatically affect the X-ray power spectrum, removing oscillations at all frequencies above 1 Hz, which effect is clearly not seen, and thus this scenario can be ruled out. The bound-free absorption in a spatially extended medium of moderate optical depth associated with the outer regions of the disc appears to be ruled out as there are a rather weak or no energy dependencies of the modulation, see L06 and Section \ref{s:period} below. (Bound-free absorption in the wind from the companion is responsible for the {\it orbital\/} modulation of the X-rays in Cyg X-1, \citealt{wen99}). On the other hand, a viable scenario is the wind/corona around the outer disc being almost fully ionized, with scattering away from the line of sight being responsible for the X-ray superorbital modulation.

\begin{figure*}
\centerline{\epsfig{file=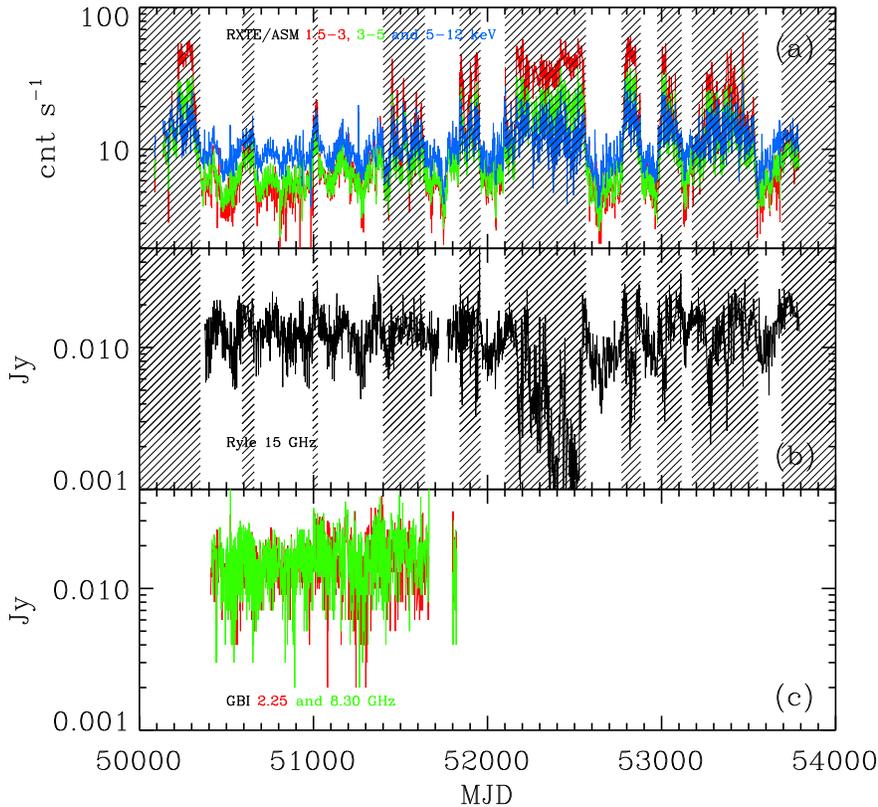,width=12cm}}
\caption{The (a) 1.5--3, 3--5 and 5--12 keV (red, greed and blue, respectively) {\it RXTE}/ASM, (b) Ryle, (c) 2.25 and 8.30 GHz (red and green, respectively) GBI light curves for the available span of the data. The shaded areas show the data not taken into account in our analysis.}
\label{f:lc}
\end{figure*}

Yet another possibility is that the X-ray emission is intrinsically anisotropic. Such a possibility was considered by \citet{brock2} who have relied on the blackbody-type anisotropy, where the flux is proportional to the projected area. However, there is overwhelming evidence that the dominant radiative process producing X-rays in the hard state of Cyg X-1 (and other black-hole binaries) is thermal Comptonization \citep[e.g.][]{p98,zg04}. Therefore, it is of interest to study models of the anisotropy of the Comptonized emission to see whether they can reproduce the observed superorbital variability.  

We note that the hot Comptonizing plasma most likely also forms the base of the jet, which is present in the hard state and radiates, at larger distances from the black hole, nonthermal synchrotron radio emission correlated with the X-rays \citep*[e.g.][]{gfp03}. However, the radiative process giving rise to the X-rays is still thermal Comptonization. Early models accounting for the radio-X-ray correlation postulated that the observed X-ray emission is due to nonthermal synchrotron emission of very energetic power-law electrons with a fine-tuned high-energy cutoff. However, that process cannot account, e.g., for the sharpness of the observed cutoffs \citep{zdz03}. Then, recent X-ray jet models turned to thermal Comptonization to account for the high-energy cutoff. However, the electron temperature in that model is very high, $kT\sim 3$--4 MeV \citep*{mnw05}, yet still fitted to the cutoff observed at $\sim$100 keV. Those authors do not explain how it is done; if it is due to the 1st order scattering by the thermal electrons, very strong fine tuning in all hard-states of black-hole binaries is obviously required. Generally, assuming the jet X-ray origin also leads to a number of other conflicts with the observations 
\citep[e.g.][]{pz03,zdz03,zdz04,mac05,yuan07}.

The synchrotron, radio, emission of the jet may be isotropic in the comoving frame (in the presence of a tangled magnetic field). However, if the jet bulk motion is at least mildly relativistic, the observed flux will depend on the jet angle, and will thus change with the jet precession. Indeed, \citet{st01} find that the absence of an observable counter jet requires such a velocity for the extended part of the jet. This constraint combined with one from the lack of radio--X-ray correlations on short time scales \citep{gl04} leads to the estimate on the jet velocity of $\sim\! (0.5$--$0.7)c$. If the jet precesses together with the disc, the radio emission will be modulated with the precession period. On the other hand, we have no information on the velocity in the core of the jet, which can be in principle much lower. In that case, the process responsible for the radio superorbital modulation may be free-free absorption in the wind from the companion star, which will depend on the direction the jet is inclined. This may provide an alternative explanation of the superorbital modulation of the radio emission \citep{sz07}. Furthermore, both the Doppler beaming and precession-dependent absorption may take place in Cyg X-1.

Here, we study the precession physical scenarios leading to superorbital modulations in a systematic way. In general, any emission anisotropic (in the rest frame of the system) with respect to the disc/jet axis will be observed to be modulated when the direction of that axis changes.

\section{The data}
\label{s:data}

We use the X-ray dwell data from the All-Sky Monitor (ASM) aboard {\it Rossi X-ray Timing Explorer\/} (\xte; \citealt*{brs93,lev96}). We also use {\it Compton Gamma Ray Observatory\/} BATSE data in the 20--100 keV and 100--300 keV energy ranges, the 15-GHz radio data from the Ryle Telescope of the Mullard Radio Astronomy Observatory, and the 2.25 and 8.30 GHz data from the Green Bank Interferometer (GBI) of the National Radio Astronomy Observatory, Green Bank, WV, USA. See L06 for a detailed description of those data sets. Compared to the analysis of L06, we also include more recent ASM and Ryle data. We study the \xte/ASM data for MJD 50087--53789 (1995 May 1--2006 February 23), the BATSE data for MJD 48371--51686 (1991 April 25--2000 May 22), the 15 GHz Ryle data for MJD 50377--53791 (1996 October 10--2006 February 25), and the GBI data for MJD 50409--51823 (1996 November 22--2000 October 6). Hereafter, we refer to the \xte/ASM channels of 1.5--3, 3--5 and 5--12 keV as the ASM A, B and C, respectively, and the BATSE 20--100 keV and 100--300 keV data as the BATSE A and B, respectively.

Cyg X-1 is a highly variable source. Therefore, in order to accurately analyze its superorbital variability it is preferable to use observations affected in the least way by its aperiodic variability. Thus, we study only the hard state, in which Cyg X-1 is for majority of the time. We define it following \citet{z02}, requiring the average photon spectral index derived from the \xte/ASM fluxes to be $< 2.1$. In addition, we exclude hard-state intervals with high X-ray variability, following the criterion used by L06. Namely, we define MJD 50660--50990 as our reference interval. We then include only those 30-d intervals of the ASM data where $< 40$ per cent of points exceed by $4\sigma$ the average flux in the reference interval. This has resulted in removal of the following time intervals: MJD $<50350$, 50590--50660, 50995--51025, 51400--51640, 51840--51960, 52100--52565, 52770--52880, 52975--53115, 53174--53554, $> 53690$. We apply the above selection to \xte/ASM and Ryle data sets, as shown in Fig.\ \ref{f:lc}. 

On the other hand, the GBI data cover a relatively short time interval, also shown in Fig.\ \ref{f:lc}. That interval is entirely in the hard state. Therefore, we have not applied any additional screening criteria to those data except the removal of the data marked as bad. Then, for the BATSE data, we exclude the periods of the soft state of 1994 and 1996, i.e., MJD 49250--49440 and MJD 50230--50307 (as in L06).
 
\begin{figure}
\centerline{\epsfig{file=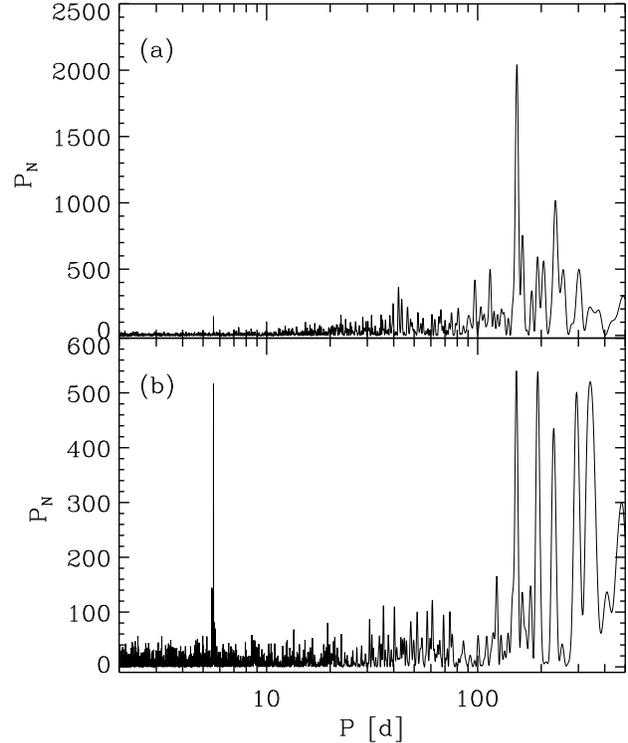,width=8.5cm}}
\caption{The Lomb-Scargle periodograms for (a) the ASM C and (b) the Ryle data (before prewhitening; corresponding periodograms after prewhitening are shown by L06).
}
\label{f:lomb}
\end{figure}

\section{Periodic modulations in the data}
\label{s:period}

The currently most comprehensive analysis of the orbital and superorbital modulations of Cyg X-1 has been done by L06. In particular, they found the superorbital period of $\sim$150 d consistent with constant in all the available data since $\sim$1976. Furthermore, they found the phase of the superorbital also compatible with constant (see fig.\ 7 and table 3 in L06).

Here, we use the Lomb-Scargle method \citep{lomb,sca} to characterize  the periodicities in the data analyzed by us. The periodograms for the ASM C and Ryle data are shown on Fig.\ \ref{f:lomb}. We clearly see the peaks corresponding to 5.6 d and $\sim$150 d for both data sets. In addition, we see strong peaks corresponding to periods longer than 150 d for the Ryle data. Their origin remains unclear though they are likely to be artefacts, see L06 for discussion. 

\begin{table*}
\caption{The coefficients for the orbital modulation fitted by equation (\ref{m:prewh}). The units of $\exp\langle G\rangle$ are s$^{-1}$ for the \xte/ASM data, and Jy for the Ryle and GBI data. }
\begin{tabular}{lccccccc} 
\hline
Detector   	& $\langle G \rangle $ 	& $G_1$         	& $\phi_1$		& $G_2$ 			& $\phi_2$ 		& $G_3$ 			& $\phi_3$ \\
\hline
\xte/ASM A 	& 1.829$\pm$0.005		& 0.113$\pm$0.008 & 0.007$\pm$0.011 & 0.042$\pm$0.008 	&--0.02$\pm$0.01	& 0.017$\pm$0.008 & --0.01$\pm$0.02\\
\xte/ASM B 	& 1.825$\pm$0.004 		& 0.066$\pm$0.006 & 0.005$\pm$0.013 & 0.022$\pm$0.006 	&0.03$\pm$0.02 	& 0.008$\pm$0.006 &--0.00$\pm$0.04\\
\xte/ASM C 	& 2.188$\pm$0.003 		& 0.035$\pm$0.005 & 0.005$\pm$0.022 & 0.011$\pm$0.005 	&--0.04$\pm$0.04	& 0.007$\pm$0.005	& --0.02$\pm$0.04\\
Ryle  		& --4.429$\pm$0.007 	& 0.167$\pm$0.011 & 0.14$\pm$0.01   & 0.035$\pm$0.010 	&0.16$\pm$0.02	& 0.017$\pm$0.010 	&0.15$\pm$0.03 \\
GBI (8.30 GHz) & --4.122$\pm$0.011	& 0.064$\pm$0.015 & 0.17$\pm$0.04  & --			& --			& --			& --		\\
GBI (2.25 GHz) & --4.187$\pm$0.010	& 0.024$\pm$0.014& 0.4  $\pm$0.1   & --			& --			& --			& --		\\
\hline
\end{tabular}
\label{t:orbmod}
\end{table*}

\begin{table*}
\caption{The periods and amplitudes of the superorbital modulation in the analyzed data. The amplitudes have been calculated using the average superorbital period of L06, 151.43 d. The errors are $1\sigma$. The second set of the $A$ values gives the modulation amplitudes obtained by L06.}
\begin{tabular}{llccc}
\hline
Detector   & Energy     & $P$ [d]         & $A$              & $A$ (L06) \\
\hline 
\xte/ASM A & 1.5--3 keV	& $153.5 \pm 1.5$ & $0.120 \pm 0.005$ & $0.133 \pm 0.005$\\
\xte/ASM B & 3--5 keV 	& $153.5 \pm 1.5$ & $0.105 \pm 0.003$ & $0.114 \pm 0.003$\\
\xte/ASM C & 5--12 keV 	& $153.5 \pm 1.3$ & $0.103 \pm 0.003$ & $0.114 \pm 0.003$\\
BATSE A & 20--100 keV & $151.0 \pm 1.1$ & $0.083 \pm 0.001$ & $0.081 \pm 0.001$ \\
BATSE B & 100--300 keV & $152.0 \pm 3.0$ & $0.075 \pm 0.002$ & $0.070 \pm  0.001$\\
Ryle    & 15 GHz 	& $152.7 \pm 1.4$ & $0.140 \pm 0.005$ & $0.105 \pm 0.002$\\
GBI& 8.30 GHz 	& $148.8 \pm 3.1$ & $0.116 \pm 0.004$ & $0.123 \pm 0.003$\\
GBI& 2.25 GHz 	& $150.7 \pm 3.5$ & $0.122 \pm 0.004$ & $0.107 \pm 0.003$\\
\hline
\end{tabular}
\label{t:ampl}
\end{table*}

We first consider the orbital modulation (caused by absorption in the wind, \citealt*{wen99,bfp02,sz07}; see the latter paper for corrections to \citealt{bfp02}). For it, we use the spectroscopic ephemeris \citep{brock, lasala98}
\begin{equation}
\mathrm{min[MJD]}=50234.79 + 5.599829 E,
\end{equation}
where $E$ is an integer. At those times, the companion star is in front of the X-ray source. We then fit the light curves folded over the orbital period and averaged within each of 20 phase bins. We use then logarithm, $G=\ln F$, of the photon or energy fluxes or count rates, $F$, for fitting with the sum of harmonics, \begin{equation}
\label{m:prewh}
G_\mathrm{mod}(\phi)=\langle G \rangle - \sum_{k=1}^{N} G_k\cos[2\upi k(\phi-\phi_k)],
\label{eq:harmonics}
\end{equation}
where $\phi$ is the 0--1 phase, $\langle G \rangle$ is the (fitted) average value of the logarithm of the flux, and $G_k$ and $\phi_k$ are the amplitude and the offset phase, respectively, for the $k$-th harmonic. We adopt the convention that $G_k>0$ and $-0.5<k\phi_k\leq 0.5$. The obtained parameters for $N=3$ for 
the \xte/ASM and Ryle data and $N=1$ for the GBI data (see L06) are given in Table \ref{t:orbmod}. We do not consider here the orbital modulation for the BATSE data since it is very weak, $\la 3$ per cent (L06).

We note that assigning errors to the folded and averaged fluxes is not a unique procedure. Here, we first divide the light curve into time bins of the 1/20 of the period (orbital or superorbital) and obtain the average of the flux, $F_{ij}$, where $i$ is the number of the phase bin and $j$ is the number of the time bin contributing to the $i$-th phase bin, weighted by the inverse squares of their measurement error. In this way, we avoid any contribution to our folded/averaged light curves from the source variability on time scales shorter than that corresponding to the length of our chosen phase bin. The resulting measurement uncertainties on $F_{ij}$ are rather small, and we neglect them. Then, we consider the average and standard deviation of all time bins contributing to a given phase bin. Here, we want to take into account the actual aperiodic variability of the source, not just the measurement errors. Therefore, we calculate the unweighted average within the $i$-th phase bin, $F_i = \left(\sum_{j=1}^{N_i} F_{ij}\right)/N_i$, where $N_i$ is the number of time bins contributing to the $i$-th phase bin, and the corresponding rms standard deviation, $\sigma_i$, given by $\sigma_i^2= \left[\sum_{j=1}^{N_i} (F_i-F_{ij})^2\right]/ [N_i(N_i-1)]$. Then, we take $G_i=\ln F_i$ and fit with equation (\ref{eq:harmonics}). 

Subsequently, we prewhiten the logarithmic light curves by subtracting from them the orbital variability fitted above, 
\begin{equation}
G'(t)=G(t)-G_\mathrm{mod}(t)+\langle G \rangle.
\label{eq:prewhiten}
\end{equation} 
We now again apply the Lomb-Scargle method to the prewhitened data. (We do not prewhiten the BATSE data, as their orbital modulation is very small, see above.) The obtained main ($\sim$150-d) superorbital periods are presented in Table \ref{t:ampl}. The given $1\sigma$ errors have been estimated using the method described in section 4.4 of \citet{sc91}. The superorbital period of Cyg X-1 was determined by L06 using the weighted average, 151.43 d, of periods determined for a large number of data sets. For our data alone we find $152.4\pm0.6$ d, and including the {\it Ginga\/} and {\it Ariel 5\/} data from L06, we obtain a similar value, $P = 152.0\pm 0.4$ d ($1\sigma$ error), consistent with that of L06. Thus, for consistency with L06, we hereafter use their ephemeris,
\begin{equation}
\mathrm{min[MJD]} = 50514.59 + 151.43 E.
\label{eq:so_ephemeris}
\end{equation}

\begin{figure}
\centerline{\epsfig{file=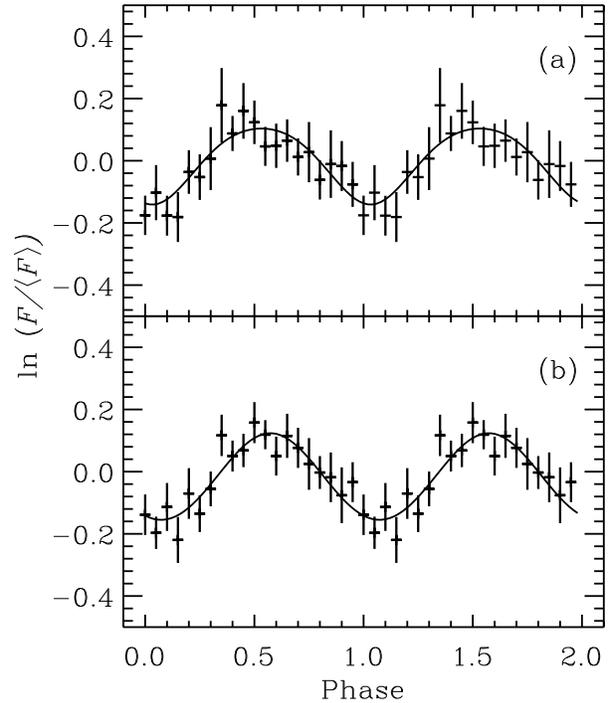,width=8.5cm}}
\caption{The superorbital phase diagrams for two of our data sets, showing the light curves folded over the superorbital ephemeris, equation (\ref{eq:so_ephemeris}), and averaged within each phase bin. (a) The ASM A X-ray data modelled by thermal Comptonization in the slab geometry. The best fit corresponds to $i=32\degr$ and $\delta=18\degr$ (see Section \ref{s:compton}, Fig.\ \ref{f:models}). The Ryle radio data fitted with the jet model (see Section \ref{s:analytic} and Fig.\ \ref{f:contour}), with the assumed values of $i=40\degr$ and $\beta=0.4$, which yield the best-fit of $\delta=11\degr$. }
\label{f:sophase}
\end{figure}

\begin{figure}
\centerline{\epsfig{file=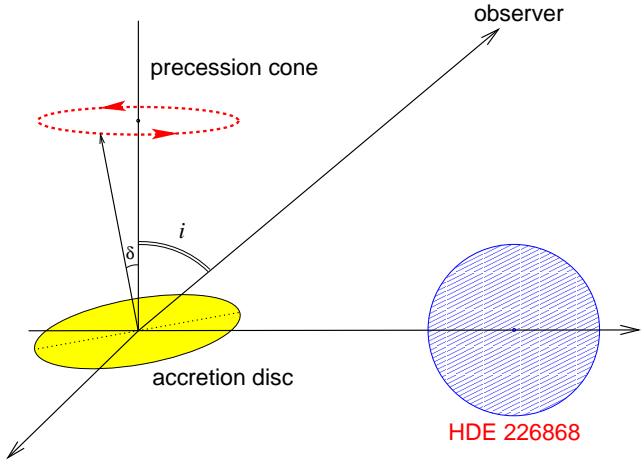,width=8.5cm}}
\caption{The assumed geometry of Cyg X-1 at the zero orbital phase (i.e., at the superiour conjunction of the X-ray source).}
\label{f:geom}
\end{figure}

We then fold and average within each of 20 superorbital phase bins the prewhitened data, and fit them using equation (\ref{eq:harmonics}) with $N=1$. The errors used in the fit are calculated in the way described above. The fits give us the amplitude of the variability, which we define as, \begin{equation}
A=\frac{F_\mathrm{max}-F_\mathrm{min}}{F_\mathrm{max}+F_\mathrm{min}} ={R-1\over R+1},
\label{eq:amplitude}
\end{equation} 
where $F_{\rm max}$ and $F_{\rm min}$ are, respectively, the maximum and minimum fitted fluxes, which ratio is $R\equiv F_{\rm max}/F_{\rm min}$, and $R= (1+A)/(1-A)$. For our harmonic fit with $N=1$, $F_\mathrm{max, min}=\exp(\langle G \rangle \pm G_1)$. The obtained values of $A$ are listed in Table \ref{t:ampl}. We see that the amplitudes of the superorbital modulation in both the ASM and radio data are almost independent of the energy, confirming the previous result of L06. On the other hand, we confirm the result of L06 that the modulation amplitudes in the 20--300 keV energy range are significantly smaller than those in the 1.5--12 keV range. Two of the resulting superorbital phase diagrams are shown in Fig.\ \ref{f:sophase}.

We point out that the obtained amplitudes are somewhat affected by systematic uncertainties, in particular those related to the selection criteria. We have tested that indeed changing the criteria used to choose the fitted parts of the light curves results in some changes of the values of $A$. To illustrate this effect, Table \ref{t:ampl} also gives the values of $A$ of L06, who used both somewhat different selection criteria and different prewhitening method (namely, prewhitening the data with all other significant modulations, not only the orbital one). We can see that their values differ from ours by typically 10 per cent, which can be taken as the systematic uncertainty of the values of $A$. The largest difference is for the Ryle data, which appears to result from not applying the selection criteria obtained from the ASM data to the Ryle data in L06. Also, we note that L06 treated each modulation as due to absorption, which lead to increases of the average fluxes after each prewhitening. In our method, we preserve the average fluxes, see equation (\ref{eq:prewhiten}).

\section{Flux variation from precession}
\label{s:precession}

Let us consider a precessing slab model with a constant tilt angle, $\delta$, see Fig.\ \ref{f:geom}, and a constant angular velocity $\omega = 2\upi P^{-1}$, where $P$ is the superorbital period. Then, the superorbital phase is proportional to time within each cycle, e.g., $\phi_{\rm so}=t/P$ for $t\leq P$. Note that the precession is independent of the orbital motion of the system, i.e., the disc inclination in an inertial frame does not directly depend on the orbital phase. The precession may also include any structure attached to the disc, e.g., a corona and/or a jet. The angle, $\psi$, between the normal to the disc and the direction towards the observer is given by
\begin{equation}
\cos\psi = |\cos\delta\cos i - \sin\delta \sin i\cos (2\upi \phi_{\rm so})|,
\label{eq:psi}
\end{equation}
where $i$ is the orbital inclination to the line of sight. Here, $\phi_{\rm so}=0$ corresponds to the disc being most inclined with respect to the observer. The emissivity of the disc (and any additional structure, e.g., a jet) integrated over its surface and over the orbital period can be only a function of the direction with respect to the normal, i.e., $\psi$. The value of $\psi$ varies between $i+\delta$ at $\phi_{\rm so}=0$ and $|i-\delta|$ at $\phi_{\rm so}=0.5$. We note that since no secondary maxima are observed in the X-ray folded superorbital light curves (Fig.\ \ref{f:sophase}), we see only one side of the disc, i.e., always $\psi<90\degr$, implying $i+\delta<90\degr$. Otherwise, we would see at some moment only the disc edge and then its opposite side. Note also that equation (\ref{eq:psi}) is symmetric with respect to an exchange of $i$ and $\delta$. Thus, for any configuration with $i>\delta$ there is an equivalent one with $\delta>i$ but with the two values exchanged. Thus, we will consider only the case of $i>\delta$, keeping in mind the existence of the symmetry.

\subsection{Analytical models}
\label{s:analytic}

If the emission from the precessing structure is anisotropic, flux variations over the precession period take place. We consider first a few simple models. Here, $0\degr \leq \psi\leq 90\degr$. 

(a) $F(\psi)= C \cos\psi$. This case corresponds to a slab/disc emitting with a constant specific intensity, i.e., blackbody-like.

(b) $F(\psi)=C\cos\psi(1+a\cos\psi)$. A more general case, where $a$ is an anisotropy parameter parameterizing departures from the blackbody. Here, $a=0$ corresponds to the constant specific intensity (blackbody) case, $a=2$ is for Chandrasekhar-Sobolev optically-thick electron scattering atmosphere \citep{cha60,sob63}, and some values in the $-1<a<0$ range correspond to Comptonization in a slab with a moderate optical depth (\citealt{st85,vp04}; see also Section \ref{s:compton} below). Values of $a<-0.5$ correspond to the Thomson optical depth across the slab $\tau \la 1$. 

(c) $F(\psi)=C[\gamma(1-\beta\cos\psi)]^{-(1+\Gamma)}$, where $F$ is the energy flux. This law corresponds to the emissivity pattern of a steady jet or disc outflow when the retardation effect is absent \citep[see e.g.][]{rl79,s97}.  Here $\beta=v/c$ is the bulk velocity in units of $c$, $\gamma= (1-\beta^2)^{-1/2}$, and $\Gamma$ is the observed photon spectral index. Note that if this model is applied to the jet case, the counter jet may be visible as well (unless obscured by the disc and/or free-free absorbed in the wind, \citealt{sz07}). In the counter jet is visible, a second term with $\beta$ replaced by $-\beta$ should be added to the above expression for $F(\psi)$.

(d) $F(\psi)=C {\rm e}^{-\tau/\cos\psi}$, where $\tau$ is the optical depth. This law corresponds to anisotropy caused by an absorber with the same symmetry as the disc.

\begin{figure*}
\centerline{\epsfig{file=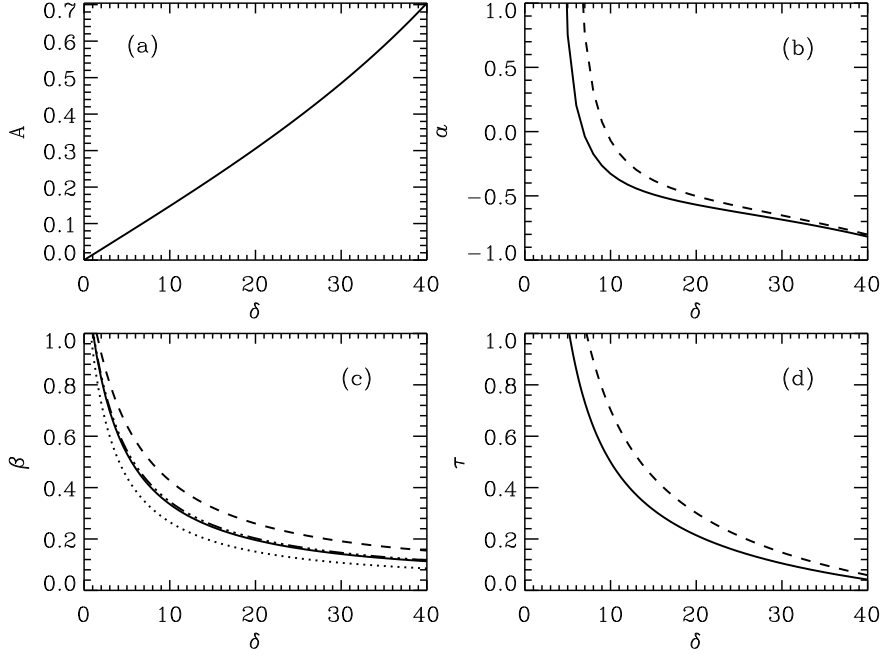 ,width=13cm}}
\caption{The relations between amplitude $A$, the precession angle, $\delta$, and other 
parameters of our analytical models for $i=40\degr$. The panels (a--d) correspond to the models (a--d) 
described in Section \ref{s:analytic}. (a) The dependence (\ref{m:modela}) between $A$ and $\delta$. (b-d) The dependencies between model parameters and the precession angle $\delta$. The assumed values of the variability amplitude of $A = 0.10$ and 0.14 are shown by the solid and dashed curves, respectively. In the case (c), $\Gamma=1$ was used. We also show the results at $\Gamma=1.7$ for the above two values of $A$ by the dotted and dash-dotted curves, respectively.}
\label{f:analytic}
\end{figure*}

Using these models, we relate either the amplitude, $A$, or the flux ratio, $R$, to the model parameters. We assume $\delta<90\degr -i$, in which case the global flux minimum corresponds to $\psi=i+\delta$. For the case (a), we have, 
\begin{equation}
\label{m:modela}
A=\tan\delta\,\tan i.
\end{equation}
In the case (b), 
\begin{equation}
A=\frac{\left( 1 + 
   2a\cos \delta\cos i
   \right) \sin \delta
  \sin i}{\cos \delta
   \cos i
   \left( 1 + 
    a\cos \delta\cos i
    \right) + 
  a\ {\sin^2 \delta} \ 
   {\sin^2 i}},
\label{eq:b1}
\end{equation}
and
\begin{equation}
a=\frac{\sin \delta\sin i-A
    \cos \delta\cos i}{
A {\cos^2 \delta} \ 
   {\cos^2 i} - 
  2\cos \delta\cos i
   \sin \delta\sin i + A
   {\sin^2 \delta} \
   {\sin^2 i}}.
\label{eq:b2}
\end{equation}
Note that in equations (\ref{eq:b1}--\ref{eq:b2}) we have assumed that the flux monotonously decreases with increasing disc viewing angle, $\psi$. This is generally the case for $a\geq -0.5$. If $a<-0.5$, the above formulae apply only
in the range of $\psi \geq \arccos(-1/2a)$. Otherwise, the maximum flux is achieved either at $\psi =\arccos(-1/2a)$ or at the maximum value of $\psi$ of the precession (i.e., $i+\delta$). In the former case, there will be two local minima of the flux, and in the latter, the minimum flux would be achieved at the minimum value of $\psi$. Both those cases can be ruled out for Cyg X-1. First, only one minimum per superorbital period is seen, and second, the X-ray superorbital modulation is in phase with the radio one (L06), and it is highly unlikely that our case (b) with $a<-0.5$ would apply to the jet radio emission. Therefore, we hereafter assume that that the flux monotonously decreases with 
increasing $\psi$ and equations (\ref{eq:b1}--\ref{eq:b2}) are applicable.
If anisotropy is interpreted as resulting from Compton scattering is a slab, the above constraints on $a$ also mean that the slab Thomson optical depth is $\tau \ga 1$, which is consistent with the estimates of $\tau $ from the X-ray spectra (see Section \ref{s:compton}). 

In the case (c), we can obtain formulae relating the flux ratio, $R$, to $\beta$ (neglecting hereafter the counter jet emission),
\begin{eqnarray}
\lefteqn{
R=\left[ 1-\beta \cos(i+\delta)\over 1-\beta \cos(i-\delta) \right]^{1+\Gamma},}\\
\lefteqn{
\beta={R^{1/(1+\Gamma)}-1\over R^{1/(1+\Gamma)} \cos(i-\delta)- \cos(i+\delta)}}
\end{eqnarray}
\citep[see also][]{g89}.
Note that for given $R$ and $i$, there is a low limit on $\delta$ (as  $\beta<1$). The minimum $\beta$ required to explain an observed flux ratio at a given inclination is achieved when the precession angle is given by 
\begin{equation}
\tan\delta ={R^{1/(1+\Gamma)}+1\over R^{1/(1+\Gamma)}-1} \tan i \, . 
\end{equation}
Note that this value of $\delta$ is generally $>i$, i.e., the interchange symmetry between $i$ and $\delta$ is no longer present here.

In the case (d), we find,
\begin{eqnarray}
\lefteqn{
R=\exp\left\{\tau[\sec(i+\delta)-\sec(i-\delta)]\right\} , }\\
\lefteqn{
\tau={\ln R\over \sec(i+\delta)-\sec(i-\delta)}.}
\end{eqnarray}

\begin{figure*}
\centerline{\epsfig{file=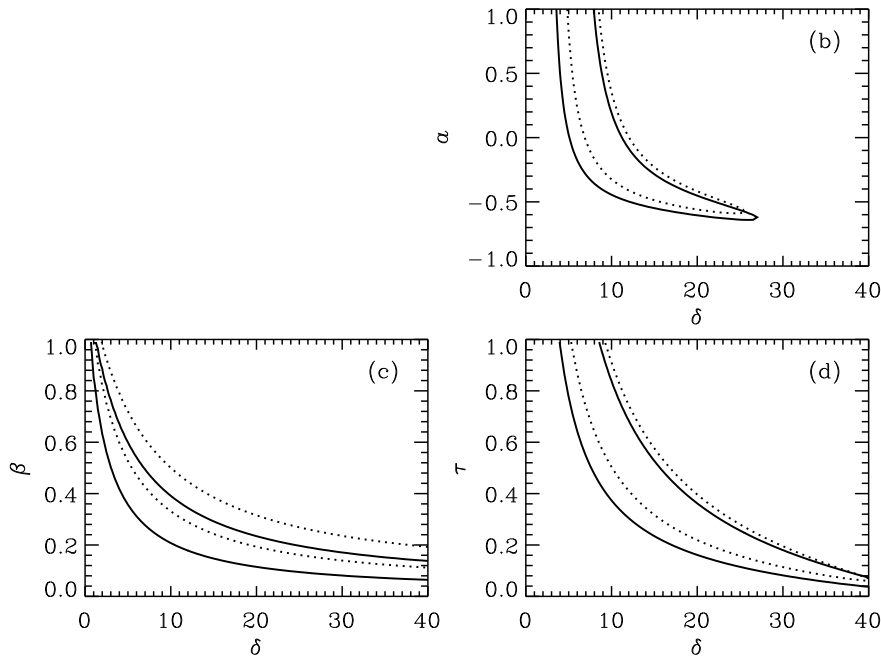 ,width=13cm}}
\caption{The 90\% confidence contours for the models (b--d) (marked on the panels) fitted to the observed superorbital variability profiles (Section \ref{s:period}) at the assumed $i=40\degr$. In the case (c), we use $\Gamma=1.7$ and 1 for the fits to the X-ray and radio data, respectively. The solid and dotted contours are for the ASM A data and the Ryle data, respectively. The ASM B, C and GBI data yield similar respective contours.}
\label{f:contour}
\end{figure*}

Various authors have obtained different constraints on the inclination of Cyg X-1, e.g., \citet*{gies86}, \citet*{wen99}, \citet*{abu04} and \citet{zi05}
got $i=28\degr$--$39\degr$, $10\degr$--$40\degr$, $31\degr$--$44\degr$, and 
$23\degr$--$38\degr$, respectively. Hereafter in this Section, we assume $i=40\degr$. Relationships between $\delta$ and other model parameters are shown in Fig.\ \ref{f:analytic}. 

We then fit the anisotropy laws (a--d) to the observed profiles of the superorbital variability (Section \ref{s:period}). The (blackbody-like) model (a) yields $\delta= 8\fdg2 \pm 0\fdg5$, $7\fdg2 \pm 0\fdg5$, $7\fdg0 \pm 0\fdg5$,  $9\fdg5 \pm 1\fdg5$, $8\fdg4 \pm 1\fdg7$ and $\delta= 8\fdg0 \pm 1\fdg6$ for the ASM A, B and C bands, Ryle and GBI 2.25 and 8.30 GHz data, respectively. For the models (b--d), the correlations between the parameters (see Fig.\ \ref{f:analytic}) lead to elongated error contours, which we show in Fig.\ \ref{f:contour}. The values of the statistic $\chi^2_\nu= \chi^2/\mbox{dof}$ for the best fits are as follows for the ASM A, B and C bands, the 15 GHz Ryle data and the 2.25 and 8.30 GHz GBI data, respectively. For the model (a): $\chi^2_\nu =(8/17$, 11/17, 14/17, 10/17, 11/17, 21/17), (b): (8/16, 11/16, 14/16, 10/16, 11/16, 17/16), (c): (9/16, 11/16, 13/16, 11/16, 12/16, 22/16), (d): (8/16, 11/16, 14/16, 10/16, 11/16, 17/16).  Thus, we see that we cannot distinguish between the models based on the fit quality. 

For the model (c), we use $\Gamma=1.7$ (the average X-ray power-law index of the hard state of Cyg X-1) for the ASM data, and $\Gamma=1$ (corresponding to the observed 2.2--220 GHz radio emission of Cyg X-1 in the hard state, \citealt{f00}) for the Ryle data. In the latter case, we do not include emission of the counter jet. The results for those two cases are consistent with a coronal outflow at $\beta \simeq 0.3$ explaining the X-ray Compton reflection strength \citep*{mbp01}, and the inferred radio jet velocity of $\beta\simeq 0.5$--0.7 in Cyg X-1 \citep{st01,gl04}, respectively. The model (c) fitted to the Ryle data superorbital phase diagram is shown in Fig.\ \ref{f:sophase}(b).

We point out that the jet in Cyg X-1 is embedded in the stellar wind from the companion, which causes the orbital modulation of the radio emission via free-free absorption \citep{sz07}. Thus, a jet precession, which changes the path of the jet during the course of the orbital motion, will also change the average optical depth in the wind along the line of sight. This will give rise to a superorbital modulation in addition to that caused by the jet beaming.

\subsection{Anisotropy of thermal Comptonization}
\label{s:compton}

\begin{figure}
\begin{center}
\leavevmode \epsfxsize=8.5cm \epsfbox{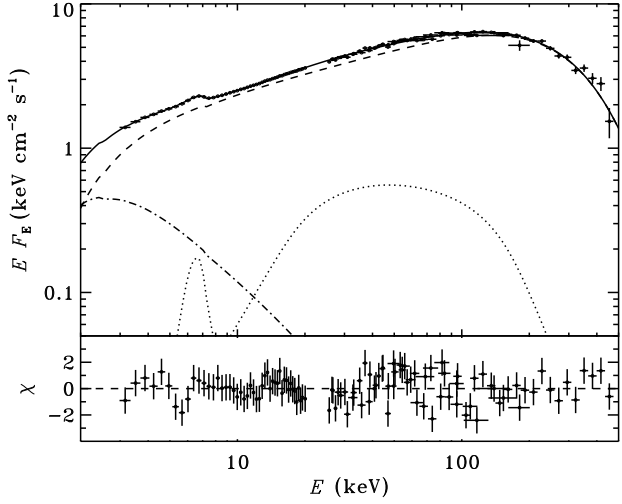} 
\end{center}
\caption{An example of the broad-band X-ray spectrum of Cyg X-1 in the hard state (data points), see Section \ref{s:compton}. The data are fitted by a thermal Comptonization and reflection model. The solid, dashed, dot-dashed and dotted curves represent the total model, the main Comptonization component, the soft excess (also fitted by Comptonization), and the reflection component together with an Fe K$\alpha$ fluorescence line, respectively. At low energies, the spectrum is absorbed by interstellar and circumstellar media. The bottom panel shows the residuals to the fit. 
\label{f:spectrum} }
\end{figure}

\begin{figure}
\centerline{\epsfig{file=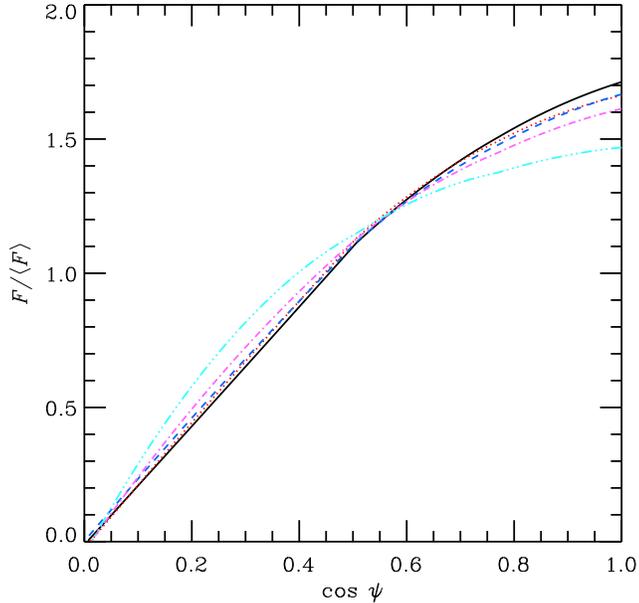,width=8.5cm}}
\caption{Anisotropy patterns of thermal Comptonization in a slab, obtained for our representative X-ray spectrum of Cyg X-1 in the hard state, Fig.\ \ref{f:spectrum}. The solid, dotted, dashed, dot-dashed, and three-dot-dashed curves give the photon flux integrated over the 1.5--3 keV, 3--5 keV, 5--12 keV, 20--100 keV, and 100--300 keV energy bands, respectively. The curves are normalized to $\langle F \rangle=\int_{0}^{1} F(\cos\psi){\rm d}\cos\psi$. The blackbody emission pattern would be represented by a diagonal straight line, whereas the anisotropy law of the case (b) of Section \ref{s:analytic} with $a=-0.4$ lies between the shown dependencies for 5--12 keV and 20--100 keV. }
\label{f:anisotropy}
\end{figure} 

\begin{figure}
\centerline{\epsfig{file=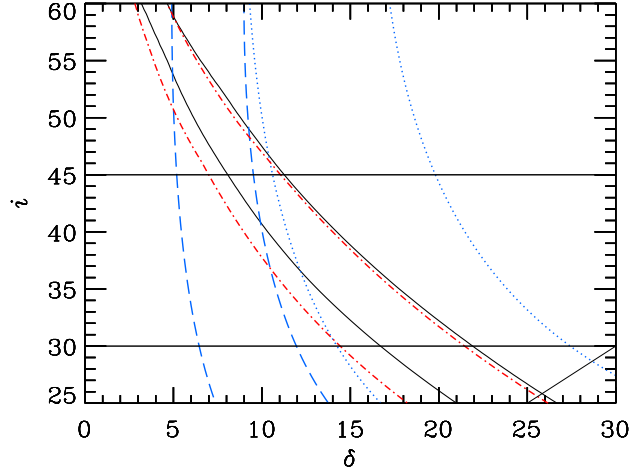 ,width=8.5cm}}
\caption{The 90 per cent confidence contours for the inclination and the precession angle. The solid and dot-dashed curves correspond to the joint fit of all three ASM data and the BATSE A data, respectively. The dotted and dashed curves show the confidence contours of the fit to the Ryle data for the jet model with $\Gamma=1$ for $\beta=0.3$ and 0.5, respectively. The two horizontal lines mark $i=30\degr$ and $45\degr$, approximately corresponding to the range of the inclination of Cyg X-1 allowed by other observational constraints. The line in the bottom right corner corresponds to $i=\delta$, below which there is a symmetric, $i<\delta$, solution (ruled out for Cyg X-1 by the $30\la i \la 45\degr$ constraint).}
\label{f:models}
\end{figure} 
 
The dominant radiative process giving rise to X-rays in the hard state of Cyg X-1 is, most likely, Comptonization of some soft seed photons by predominantly thermal electrons at the temperature of $\sim$50--100 keV \citep{gier97,p98,d01,f01,z02,mc02,i05}. This is evidenced mostly by the characteristic form of its high-energy cutoff, present also in other black-hole binaries in the hard state \citep[e.g.][]{g98,zpm98,w02,zg04}. The location of the thermally Comptonizing plasma is either a hot inner accretion flow \citep*[e.g.][]{pkr97,e98,p98,z02,yuan07} or a coronal outflow \citep{b99,mbp01}. 

Thus, we fit such a model to a representative hard-state spectrum of Cyg X-1. We choose the \xte\/ observation 10238-01-03-00 together with the \gro/OSSE observation VP 612.5 (spectrum 6 of \citealt{i05}), which gives us a broad-band energy coverage of $\sim$3--1000 keV. Our detailed model consists of the main component due to thermal Comptonization in a slab geometry calculated using the iterative scattering method of \citet{ps96} ({\sc compps} in XSPEC, \citealt{a96}). In addition, we include Compton reflection \citep{mz95}, a fluorescent Fe K$\alpha$ line, and a soft excess. We attribute the last component to thermal Comptonization as well, but in a plasma with the Compton $y$ parameter much smaller than that of the main Comptonizing plasma, and neglecting its Compton reflection \citep[as in][]{f01}. Here, $y\equiv 4\tau kT_{\rm e}/m_{\rm e}c^2$, where $\tau$ is the optical depth of the plasma, and $T_{\rm e}$ is its electron temperature. The seed photons for Comptonization are from a blackbody spectral component at the temperature of $T_{\rm bb}$, which we keep equal to 0.2 keV in the fit. Also, we assume the inclination of $\psi =40\degr$. The spectrum with the resulting fit components is shown on Fig.\ \ref{f:spectrum}. The main fit parameters are $kT_\mathrm{e}=88$ keV and $\tau=1.17$ for the main Comptonization component, the solid angle of the Compton reflector of $\Omega/2\upi =0.17$,  and $\tau=0.87$ of the soft-excess Comptonizing plasma, for which $kT_{\rm e}=20$ keV was assumed. The absorber column density is $N_{\rm H}=2.3\times 10^{22}$ $\mathrm{cm}^{-2}$. The fit statistic is good, $\chi^2_\nu=371/401$. 

We then vary the inclination of this model, and for each $\psi$ we integrate the photon flux in five energy bands, 1.5--3 keV, 3--5 keV, 5--12 keV, 20--100 keV, 100--300 keV. The resulting dependencies of $F(\psi)$ are shown in Fig.\ \ref{f:anisotropy}. We also compare the resulting functions to $F(\psi)= \cos\psi(1+a\cos\psi) /(1/2+a/3)$, i.e., the normalized case (b) of Section \ref{s:analytic}. We find $a\simeq -0.4$ as the closest overall approximation 
for the shown angular dependencies. On the other hand, the normalized blackbody dependence equals $F(\psi)=2\cos\psi$, which yields less (more) emission than the normalized thermal Comptonization dependencies at $\cos \psi \la 0.65$ ($\ga 0.65$).

We now use the dependence, $F(\psi)$, in an energy band together with $\psi(i,\delta,\phi_{\rm so})$ of equation (\ref{eq:psi}) to fit the $F(\phi_{\rm so})$ observed in that energy band, with $i$ and $\delta$ as the fitted parameters. We use here the ASM and BATSE data. (Since the BATSE data are given as energy fluxes, we use the corresponding integrated energy fluxes for them.) This yields very good fits to all the data, e.g., $\chi^2_\nu$ at the best fits to the ASM A, B, C data, and the joint fit for all three detectors equal 9/17, 11/17, 13/17 and 34/57, respectively. We have calculated the 90 per cent confidence contours for all the data, and found that they remain in excellent mutual agreement. We show the contours for the ASM joint fit and for the BATSE A data in Fig.\ \ref{f:models} by the solid and dot-dashed curves, respectively (the BATSE B contours are very similar to those of the BATSE A, except for extending to a somewhat higher value of $\delta$ at a given $i$).

This agreement provides a strong argument for the correctness of our model of anisotropy as due to thermal Comptonization in a geometry similar to that of a slab. Namely, we see in Table \ref{t:ampl} that the fractional X-ray modulation decreases monotonically with the increasing photon energy band, from 1.5 keV to 300 keV. In particular, the ASM A channel shows the amplitude higher than that of the channels B and C, and the BATSE data show the amplitudes lower than those of the ASM. This is in excellent agreement with the theoretical anisotropy patterns shown in Fig.\ \ref{f:anisotropy}. Namely, we see there that the degree of anisotropy of thermal-Compton emission, as measured by the slope of the curves, decreases monotonically with the increasing energy (for $\cos\psi\ga 0.3$, which is satisfied in all of our fits). This effect is due to the average number of scattering in the plasma increasing (thus leading to less anisotropic emission) with the photon energy. This then leads to a corresponding decrease of the relative amplitude of the precessional modulation, which is both present in the X-ray data sets (Table \ref{t:ampl}), as well it is very well fitted by our model. 

On the other hand, we see the contours are very elongated, which results from a strong correlation between $i$ and $\delta$ in the model. The $1\sigma$ contours are only slightly smaller than the shown 90 per cent ones, indicating that the minima of $\chi^2$ are very shallow. Therefore, in order to constrain $\delta$, we use the constraint of $30\degr\leq i\leq 45\degr$ (approximately corresponding to the constraints on $i$ by various authors listed in Section \ref{s:analytic}), with the boundaries of this region shown by the horizontal lines in Fig.\ \ref{f:models}. Then, we find $8\degr\la \delta\la 11\degr$ at $i=45\degr$, and $17\degr\la \delta\la 22\degr$ at $i=30\degr$. We then compare these results with the 90 per cent confidence contours for the radio emission as fitted with the jet model. We find that model is compatible with the Comptonization model for the X-ray data if $\beta=0.3$--0.5 (shown by the dotted and dashed contours for $\beta=0.3$ and 0.5, respectively), but not if $\beta =0.7$. For $\beta=0.3$--0.5 and $30\degr\leq i\leq 45\degr$, the radio data provide constraints on $\delta$ less restrictive than the X-ray data. We show the best-fit ASM A model and a model for the Ryle data on the superorbital phase diagrams, see Fig.\ \ref{f:sophase}. 

The main caveat to our results here is related to using the slab model for Comptonization. It is unlikely that the actual X-ray source in Cyg X-1 has exactly this geometry, and it is probably less anisotropic. This would yield somewhat higher precession angles than those obtained by us, but it is unlikely to affect our qualitative conclusions. Also, we have used a specific (but rather typical) single spectrum to obtain $F(\psi)$. However, though there is some spectral variability in the hard state \citep[e.g.][]{i05}, it is aperiodic and its effect will be averaged out over a number of superorbital cycles. Thus, using this assumption is unlikely to affect significantly our results. 

\section{Discussion and Conclusions}
\label{sec:concl}

We have studied a general problem of how precession can affect the observed fluxes and spectra. Apart from an obvious case of a full or partial covering of the central source by an accretion disc (e.g., in Her X-1, \citealt{l02}), the observed flux can be modulated if the emission pattern of the precessing source is anisotropic. We have studied a number of likely anisotropy patterns, due to either intrinsic anisotropy of the emission pattern in the source rest frame, or Doppler boosting of emission isotropic in the rest frame, or absorption/scattering by a corona covering the disc. We have derived formulae for the flux as a function of the precession phase for those processes. In addition, we have numerically calculated the X-ray anisotropy pattern due to thermal Comptonization and Compton reflection in a plasma with parameters characteristic to black-hole binaries in the hard state and assuming a slab geometry. 

We have then applied these results to the superorbital modulation of the 2.25--15 GHz radio and 1.5--300 keV X-ray emission of Cyg X-1 in the hard state. We find that most models make similar predictions for the precession angle, i.e., the tilt between the disc plane and the orbital plane, of the order of $\delta \sim 10\degr$--$20\degr$. The statistical quality of all the models is similar, thus, they cannot be distinguished on the basis of their $\chi^2$. The model with absorption by a disc corona yields strongly anticorrelated the optical depth and the precession angle.

Our most physical model appears to be thermal Comptonization for the X-rays and jet Doppler boosting for the radio emission. The joint constraint to the X-ray and radio data yields for $\beta=0.3$ the values of $\delta\ga 10 \degr$, $i\la 47\degr$; and for $\beta=0.5$, the values of $5\degr\la \delta\la 10\degr$, $40\degr\la i \la 60\degr$ (see Fig.\ \ref{f:models}). If we assume $30 \degr \leq i \leq 45\degr$ (from other observational constraints on the inclination, see Section \ref{s:analytic}), we obtain $8\degr \la\delta\la 22\degr$ from the X-ray data, which then implies $\beta\sim 0.3$--0.5 using the Ryle data. Thus, both the disc/jet precession angle and the jet velocity in Cyg X-1 appear relatively small. We also point out that our Comptonization model also explains and well fits the decrease of the fractional superorbital modulation amplitude with the increasing photon energy observed in the 1.5--300 keV X-ray emission of Cyg X-1. The decrease is explained by the Comptonization anisotropy decreasing with the photon energy due to the correspondingly increasing average number of scatterings suffered by a photon before leaving the source.

Finally, it is interesting to compare our results for Cyg X-1 with those for the well-known precessing jet source SS 433. In that case, the jet velocity and the precession angle are very precisely determined based on the Doppler shift of spectral lines (formed in the jets). Their values are $\beta\simeq 0.265$ and $\delta\simeq 21\degr$ (\citealt{eik01}; see \citealt{fab04} for a review). They are rather similar to those determined by us for Cyg X-1. However, the disc emission is not seen at all due to heavy obscuration, and the radio emission shows no periodic variability on either orbital or precesssional period. This appears to be related to the rather large inclination of SS 433 of $i\simeq 78\degr$, at which the disc/jet system goes through the edge-on orientation at some point of each precessional cycle. Then, the counter jet is seen at a lower inclination than the primary jet during a part of the cycle, and both of them contribute comparably to the observed flux (see Section \ref{s:analytic}). Also, free-free absorption of the radio emission is of importance in that system \citep[e.g.][]{fab04}, which introduces a major complication to predictions of the precessional variability given that details of the wind geometry are unknown. On the other hand, the precessional modulation is very significant in the X-rays \citep{wen06}, emitted by the jets.

The disc/jet precession in SS 433 is probably caused by the precession of the companion star, with the rotation axis inclined with respect to the orbital plane (see \citealt{fab04} and references therein). On the other hand, this precession mechanism is unlikely in Cyg X-1, where the orbit is strictly circular \citep{brock2}, indicating that tidal forces have also brought the stellar rotation to alignment. Then, the precession may be caused, e.g., by the gravitational force of the star acting on the inclined disc (e.g., \citealt{l98}; L06).

\section*{Acknowledgments}

AI has been supported by the Graduate School in Astronomy and Space Physics, V\"ais\"al\"a foundation and by the Russian Presidential program for support of leading scientific schools (grant NSH-784.2006.2). AAZ has been supported by the Academy of Finland exchange grant 112986, and by the Polish grants 1P03D01827, 1P03D01128 and 4T12E04727. JP was supported by the Academy of Finland grants 102181, 109122, 110792 and 111720.  We are thankful to Guy Pooley for the Ryle data, and to Pawe{\l} Lachowicz for helpful discussions. The Ryle Telescope is supported by PPARC. The Green Bank Interferometer is a facility of the National Science Foundation operated by the NRAO in support of NASA High Energy Astrophysics programs. We acknowledge the use of data obtained through the HEASARC online service provided by NASA/GSFC.

\label{lastpage}
\end{document}